# PS2: Managing the next step in the Pan-STARRS wide field survey system

William S. Burgett*[a]

[a]University of Hawaii Institute for Astronomy, 2680 Woodlawn Dr., Honolulu, HI, USA 96822

## ABSTRACT

The Panoramic Survey Telescope and Rapid Response System (Pan-STARRS) is unique among the existing or planned major ground-based optical survey systems as the only "distributed aperture" system. The concept of increasing system étendue by replicating small telescopes and digital cameras presents both management opportunities and challenges. The focus in this paper is on management lessons learned from PS1, and how those have been used to form the management plan for PS2. The management plan components emphasized here include technical development, financial and schedule planning, and critical path and risk management. Finally, the status and schedule for PS2 are presented.

## 1. INTRODUCTION

The emphasis of this paper is on the management planning for the second Pan-STARRS telescope and camera system, PS2, that will augment the already operational first system, PS1. Technical details concerning PS2 can be found in the papers by Morgan et al. and Onaka et al. presented elsewhere at this conference[1,2]. When PS2 is fully operational, the combined two-aperture system, hereafter PS1+2, will have approximately the same étendue as the Dark Energy Survey (étendue is the product of collecting area and field of view per aperture). Although documented in more detail elsewhere, e.g., reference [3], an overview of Pan-STARRS is provided in this section for the sake of completeness. In addition, some selected general considerations concerning project management, systems engineering, and risk management are included in Section 2 to help firmly establish the context in which Pan-STARRS management issues are defined and addressed. Section 3 presents issues and status updates specifically for PS2.

### 1.1 Overview of Pan-STARRS science goals and system concept

Developed by the University of Hawaii Institute for Astronomy (UH-IfA), Pan-STARRS PS1 is currently the world's most powerful astronomical survey system. Although it cannot be presented in detail here, the scientific return from PS1 is already impressive and is increasing rapidly. The gigapixel camera (GPC1) has been listed as one of the "20 Marvels of Modern Engineering" (http://www.gizmowatch.com/entry/20-marvels-of-modern-engineering/). Pan-STARRS will create the largest ever astronomical survey and database. More importantly, it will create a motion picture of regions of the sky so objects variable in space or time can be easily discovered and followed up. The 2000 Decadal Review report, "Astronomy and Astrophysics in the New Millennium" (AANM), reviewed the status and priorities for astronomy for the first decade of the 21st century, and highlighted the need for a new 6m-class survey telescope for a modern survey system utilizing state-of-the-art imaging technologies[4]. The AANM survey science goals, the same as the science goals for Pan-STARRS, are the following:

- A census of the solar system with emphasis on finding potentially hazardous asteroids and comets, and a greatly increased inventory of objects in the outer solar system,
- Detailed studies of the content and formation history of the Milky Way,
- Dark energy (DE) and dark matter (DM) from studies of the large scale structure (LSS) of the Universe,
- Cosmology from "time-domain" observations of Type Ia supernovae that provide measurements of the expansion history of the Universe.

In order to implement a system technologically capable of achieving these goals at a reasonable cost, Pan-STARRS opted for a "distributed aperture" wide-field high-resolution design concept, eventually planned to comprise four 1.8m apertures (PS-4) with 7-square degree field of view per aperture. Combining the images from smaller telescopes is done in such a way to yield the performance of a single larger telescope at a significantly reduced cost. The planned Pan-STARRS PS-4 has a somewhat smaller effective collecting area than the original ~6m AANM goal (or the 8m primary mirror size chosen by the LSST project). However, by virtue of superior image quality provided by Hawaiian sites, PS-4 could achieve the original AANM science goals. These same goals, and the importance of wide field surveys at all wavelengths from X-ray through radio, were validated again the 2010 decadal survey report, "New Worlds, New Horizons in Astronomy and Astrophysics"[5].

*burgett@ifa.hawaii.edu; phone 1 808 988-8965;   pan-starrs.ifa.hawaii.edu/public/

Other factors and design features relevant to the Pan-STARRS design:
- Capability with PS-4 to survey the entire available sky to 24th magnitude weekly,
- Partial adaptive optics utilizing state-of-the-art CCD technology developed especially for astronomical imaging, possible with small telescopes but ineffective with larger telescopes, ,
- Multiplicity of images: summing deep images that are constructed from a very large number of individually shallow images to minimize systematic effects.

The stream of image data even from PS1 (the first of four planned telescopes that will comprise PS-4) represents an order of magnitude increase over any previous astronomical survey. PS1 collects ~500-700 images per night with its 1.4 billion pixel camera resulting in 1000-1400 Gigabytes of image data per night. Catalogs of objects produced by Pan-STARRS will eventually comprise more than 90% of the objects in our Solar System greater than ~140-300 meters in diameter, at least one billion of 100 billion stars in our Milky Way galaxy, and a billion galaxies in the cosmos.

## 1.2 Overview of Pan-STARRS subsystems

The Pan-STARRS major subsystems, interfaces, and data flow are shown in Figure 1. Brief descriptions of these subsystems are given in the following.

**Telescope:** The optics configuration is Ritchey-Chretien with a 1.8-meter primary and 0.9-meter secondary together with a 3-element wide-field corrector delivering an f/4.4 beam into the detector providing a plate scale of 0.0256″ per micron. The optical design by itself yields better than 0.4″ FWHM over the entire 3° diameter FOV. The primary mirror support is pneumatic and allows active figure control. The secondary is mounted on a hexapod providing active 6-dof control. The telescope design is fully baffled - there are no direct paths from outside the field of view to the detector. It emerged during PS1 commissioning and early operation that excess sky background was caused by moonlight illuminating the baffle surfaces which then scattered light to the detector. This problem has been resolved by a combination of ribs on baffles and changes to the baffle surface material as well as using the shutter dome as a moon-shield.

**Camera:** The detector design is driven by the sub-arcsecond image atmospheric seeing provided by Hawaiian sites, and the choice of 10-micron pixels subtending 0.256″ to adequately sample the FWHM. This results in needing ~1.4 billion pixels to cover the entire 3° diameter FOV. The short exposure times needed for NEO searches in order to avoid trailing losses lead to a requirement for very short readout times. This drove the development of a massively parallel system design. The detectors are deployed as an 8×8 array of 5-cm devices, each of which is an 8×8 array of independently addressable "detector cells". The resulting 4096 cells are read out and controlled by a 512 channel high-throughput readout system. The detectors also feature orthogonal transfer (OT) charge shifting to allow on-chip tip-tilt correction to mitigate wind-shake and seeing. An operational OT mode has been demonstrated with PS1.

**Image Processing Pipeline (IPP):** The Pan-STARRS Image Processing Pipeline (IPP) is required to perform near real-time analysis of the full data stream coming from the camera and telescope, to archive the raw images and output data products, to generate instrumental response corrections, and to perform the astrometric and photometric calibration of the resulting data. The IPP performs a series of analysis steps on every individual image as well as image stacking and image differencing. The single-image analysis stage corrects for the detector response, detects and characterizes the astronomical objects in the images, and performs an initial astrometric and photometric calibration. Collections of images are stacked to improve the depth and to fill in the gaps. Image differencing, with PSF-matching, is performed between individual pairs of images, pairs of stacks, or combinations. A high-quality object analysis is performed on the deep stacks, and the repeated and overlapping observations generate improved astrometric and photometric calibrations of the images. The same process results in a high-quality astrometric and photometric reference catalog along with proper-motion and parallax measurements across the sky. The short exposure times of typically 30 – 40 seconds result in a huge data rate leading to a design requirement for a massively parallel architecture for the IPP featuring fault tolerance, task scheduling, etc.

**Observatory, Telescope, and Instrumentation plus Scheduling software (OTIS):** The PS1 Observatory, Telescope, and Instrument Software (OTIS) system is an integrated software package for operating the PS1 observatory (dome, telescope, camera, etc.). Pan-STARRS operation is unique in some ways that drive the OTIS capability and design. The observatory and telescope are operated remotely, and this requires a significant amount of auxiliary instrumentation and meteorological sensors as well as power and network monitors to access all heartbeat functions within the observatory. Another significant OTIS capability is to efficiently schedule the observations to meet the goals of the multi-faceted survey program with all of its complicated cadences and revisiting requirements ranging from 15 minutes to six months. The scheduling capability includes maintaining a record of prior observations, the image quality obtained for each collected image, and adapting the observing schedules to accommodate weather factors.

**Optical Control System:** The wide-field design requires fine control of optical elements to a precision of better than a few tens of microns. To obtain this precision, a wave-front sensing system was developed using out-of-focus images to diagnose errors in collimation and alignment. The result is a static model that is integrated into OTIS to control collimation and alignment via positioning of the secondary mirror hexapod and the primary mirror figure controllers. We have also developed an active control system using in-focus images to provide a running correction to a static model for focus (as a function of telescope orientation, temperatures, pass-band, etc.). The performance of this system has been very good.

Techniques to use out-of-focus images were also developed to measure deviations between the detector- and focal-surfaces which inevitably arise as it is difficult to generate a perfect detector surface in the lab. For PS1, the camera was removed from the telescope at one point, and the positioning of the devices was adjusted and checked with a metrology microscope given the deviations that were measured on-sky. Aside from a fairly strong deformation of the focal surface affecting a small area in the center of the field that was left uncorrected, this was found to be highly effective.

Developing these techniques was quite challenging as it required optical analysis and developing specialized image processing for analysis of the out-of-focus 'donut' images. Altogether this took approximately two years to perfect. For PS2 we will apply the same techniques, and we expect to be able to carry out the process in a few weeks, barring serious hardware problems.

**Moving Object Processing System (MOPS):** PS1 discovers asteroids via its sophisticated Moving Object Processing System software. Generally speaking, MOPS assembles nightly associations of transient detections from the IPP into larger associations, called linkages, eventually producing an "identification" from which an orbit can be computed with sufficient confidence. MOPS detection data and processing results are stored in a relational database searchable via direct database query or web-based tools. A key component of MOPS is its ability to simultaneously run a large synthetic population through the pipeline to facilitate de-biasing and population studies. Under development since 2004, MOPS is the only Pan-STARRS project-funded science client.

**Published Science Products Software (PSPS):** The Pan-STARRS PS1 PSPS subsystem is a relational database that publishes the catalogs of detections and objects to the PS1 Science Consortium (PS1SC) scientists and ultimately to the world. The database was developed at the UH-IfA in collaboration with an experienced team from the Johns Hopkins University, and leveraged important features and lessons learned from the development of the Sloan Digital Sky Survey (SDSS) database.

## 2. COMMENTS ON SELECTED TOPICS OF PAN-STARRS PROJECT MANAGEMENT

In this section, general comments are made about some issues relevant to managing scientific projects in general and Pan-STARRS in particular. Much of the material and comments are well known and "obvious" to those experienced with large scale and complex system fabrications and facility construction exemplified by a system such as Pan-STARRS. However, while the details of the management approach and experience for Pan-STARRS are well known to the Air Force Research Laboratory sponsor (AFRL), they have not been documented in the open literature, and sometimes have been subject to anecdotal comments that have not entirely (or correctly) captured the reality. Thus, in addition to providing an update of PS2 status, an important motivation for this paper is to give some transparency and insight into the broader Pan-STARRS management approach, both the philosophy and actual implementation.

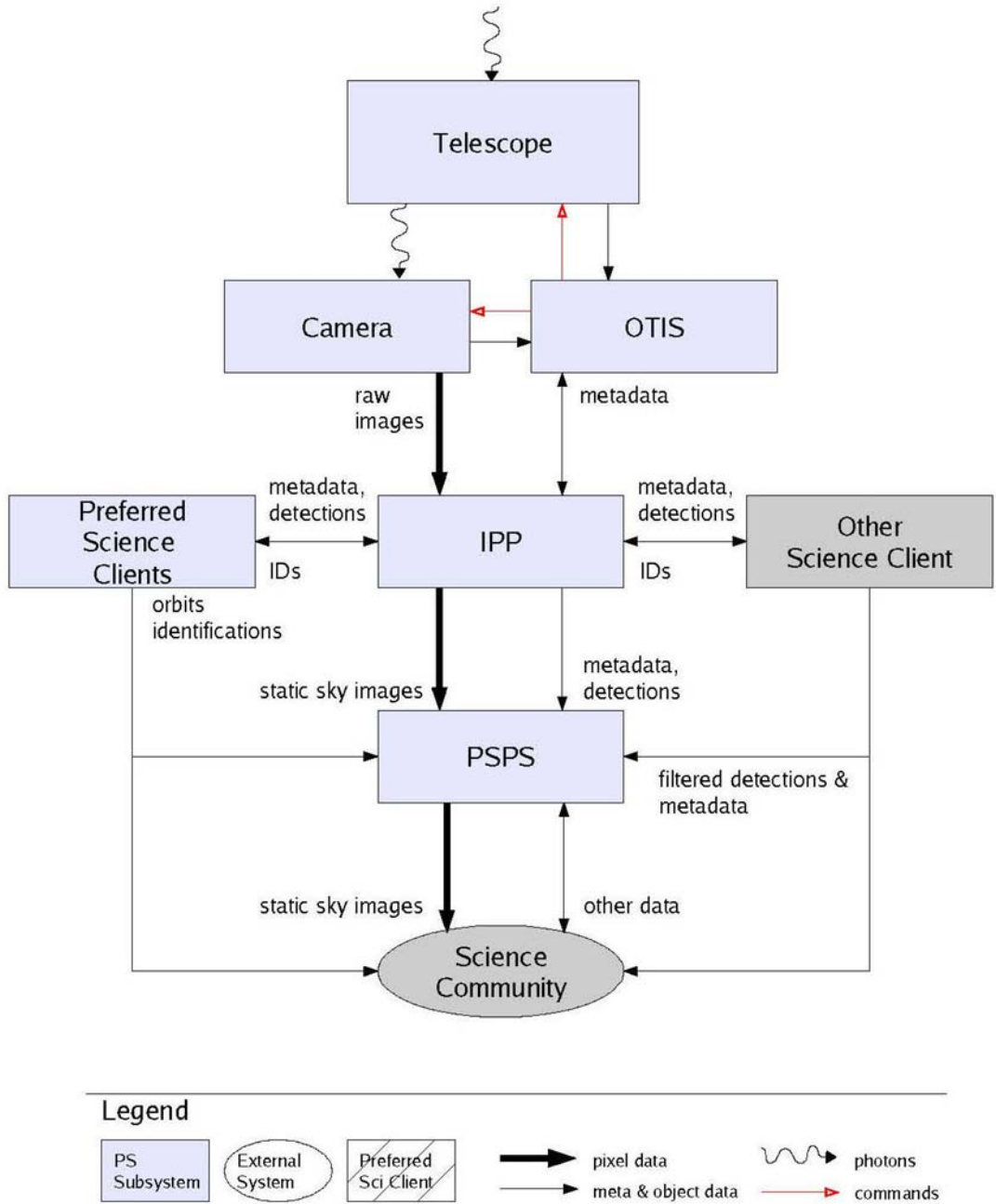

Figure 1. Block diagram showing the major subsystems and the major interfaces.

**2.1 Managing Pan-STARRS as an academically-based science project**

There have been many Pan-STARRS "lessons learned" at different levels from project planning to system design and integration and testing ("commissioning"). The intent of this section is to provide some insight into some common issues and the "Pan-STARRS experience" from a project management perspective. Experienced project managers and systems engineers will recognize the issues and comments as "par for the course".

Effective and successful project management is neither trivial nor easy. This is further complicated when a project is being executed within an academic environment involving the heavy participation by smart and capable scientists who are inexperienced in working on technologically complex and/or large projects. Such is the case with Pan-STARRS. Pan-STARRS is a technologically complex system that initially required staffing levels of a few tens of engineers, scientists, and administrative personnel at the UH-IfA plus external partners and several subcontractors. The number of major subsystems and interfaces involving largely independent groups of people, plus the technological challenges inherent within the chosen science goals and design concepts, require significant management and oversight. Achieving the commitment ("buy-in") from some project personnel working on their first project on the Pan-STARRS scale to work within a unified management plan has not always been as easy as it probably should have been. This definitely is not intended to minimize the fact that the Pan-STARRS team includes extremely talented and capable scientists and engineers, some of them among the top in their field in the world, and that without these people and their extreme dedication, Pan-STARRS would not have been successful at all.

The debate about how any project is managed sometimes obscures three fundamental facts about successful project management. The first is that any project is more likely to succeed, and be executed most efficiently, when there exists (i) a well-defined organization of authorities, roles, responsibilities, and work plan, (ii) configuration management, and (iii) good documentation. The second fact, even more fundamental, is that almost any real management plan has a chance to succeed if there is sufficient commitment ( "buy-in") by the project team from all levels above and below the project manager. Conversely, no matter how well-defined and appropriate for a project, if the project team is not committed to working within a given organizational structure and plan, chances are diminished greatly that the specified project technical performance will be realized within budget and on schedule. In other words, insufficient buy-in incurs high risk. The third fundamental fact is the management of expectations. The three principal aspects of managing expectations are providing transparency to team and sponsors, initially setting realistic goals, and maintaining the discipline to avoid requirements/scope creep. The latter two aspects are sometimes problematic for an academically-based project with a strong R&D component. The appropriate management approach is obvious: implement sufficient requirements definition, configuration management, and overall documentation. However, once again, it is absolutely necessary to have the buy-in from the project team and its leaders both above and below the project manager to enforce and to maintain this discipline. One common pushback to this general approach is the false distinction embodied in the often heard question, "Do you want something accomplished or do you want documentation?".

Finally, it is also worth noting that for good reason, a popular management tool for a technologically mature complex project with a budget in the millions of dollars (or tens of millions or more) is the use of earned value metrics (EV). Using EV provides a clear and consistent quantitative view of management performance for both the project team and its sponsors. However, EV is appropriate and useful when two conditions exist and are maintained: (1) the scope of work is clearly defined and bounded, and (2) those providing budget and schedule inputs from the bottom up recognize the criticality of providing accurate inputs. This is another aspect of the team's commitment to the management plan. Because of the first condition above, it is a fact that EV may be difficult or impossible to use for projects containing large amounts of research and development (R&D). Of course, a counter-argument can be made that even an academically-based project having an overall milestone goal of operation or production should complete all R&D before embarking on the plan to achieve operation or production. However, that is not always practical for this type of project because sponsors often award grants containing funds structured to conduct R&D that must still result in operation or production. Sponsors would probably point out that they often receive proposals from scientists promising operation or production without fully clarifying the necessary R&D to achieve the final goal. The solution here is a combination of initially managing expectations, defining the resultant scope of work, avoiding scope creep, and being willing to compromise on technical challenges in order to produce a final result. In other words, regardless of whether a system or tool such as EV is employed, it is always in everyone's best interest to never allow "the perfect to be the enemy of the good enough".

## 2.2 Summary of risk assessment principles

It is understood that risk management (assessment, control, and mitigation) is a primary factor in determining the capability of maintaining any proposed budget and schedule. Therefore, more detail is given here concerning these issues including the clear definition of terms, methodology, and reasoning in order to provide a clear explanation of risk, risk analysis, risk control and mitigation, and overall risk management. We do not mean to appear unnecessarily pedantic, but we believe that given the importance of this management component, it is better to err on the side of providing more detail than is perhaps truly necessary rather than not providing enough detail that leaves critical questions unanswered.

Our formal definition of risk is probability-based in that the higher the probability of occurrence that an event or combination of events can adversely impact technical performance, budget, or schedule beyond the availability contingency to mitigate the impact, then the higher the risk. The three associated probability ranges we use are Low probabilities from 0.00 – ≈0.15, Medium probabilities from ≈0.15 – ≈0.40, and High probabilities from ≈0.40 – 1.00. The probability ranges chosen are generally considered to be project-specific and somewhat arbitrary, but in most cases will not vary greatly from the ranges chosen here for projects similar to Pan-STARRS. The probabilities are then combined with estimates of the adverse consequence of any single event or sequence of events, e.g., some events have relatively low adverse consequences regardless of the probability of occurrence. The adverse consequence – probability of occurrence combinations are what we label as "impact risk," or simply "risk". We use three overall risk levels as sufficiently fine-grained for an overall estimate of the PS2 program plan: Low, Medium, and High. Then, a high consequence, high probability event has a high impact, a medium consequence, low probability has a less than medium impact, etc. With the low, medium, high labels applied to a consequence-probability matrix, one defines the various impact levels as shown in Figure 2. However, as mentioned above, for the risk assessments used for Pan-STARRS, we find it an adequate working approximation to classify the (LM, LL, and ML) impacts as Low, the (LH, MM, and HL) impacts as Medium, and the (MH, HH, and HM) impacts as High. The basic goal of risk management then becomes to try, as much as possible, to lower to Low those work blocks or individual task elements that are initially assessed at Medium or High impact. The three risk levels apply separately to each of the Technical, Budget, Schedule categories, but of course are often closely correlated.

Regardless of the formal nature of the definitions, risk assessment still contains subjective elements. To minimize the subjectivity, it is usually a good idea to involve multiple experienced engineers and managers in risk assessment whenever possible. In large projects, this might be formalized by a Risk Review Board reviewing a risk database. Regardless of the formality invoked by a specific project, it is obvious that effective risk control relies on frequent monitoring and updates to previous risk assessments. For example, if a task is estimated to require 6 months of actual time, but there are 24 months available in the master schedule to complete the work, an initial probability of schedule slip is Low, but if after 22 months the task has been continually delayed so that 2 months are still required to complete it, then the initial Low probability of a schedule slip has obviously changed to High. The adverse consequences can also change from initial estimates, e.g., in budgeting for a task if new information increases initial costs or if initial contingency has been expended in other areas reducing available contingency for the task under consideration.

| Impact Risk Matrix | | | |
|---|---|---|---|
| High | LH | MH | HH |
| Med | LM | MM | HM |
| Low | LL | ML | HL |
| Consequence ↑ Prob. of Occur. → | Low | Med | High |

Figure 2. The impact risk matrix showing the levels of impact of an event defined as the adverse consequence if the event occurs weighted by its probability of occurrence.

The integration and testing of any complex system necessarily requires a large number of individual, often related, tasks. Pan-STARRS is no exception. The key to successful risk control and mitigation during the integration and test of such systems is to thoroughly understand in advance the full extent of the required tasks, the connections and dependencies, and to create contingency plans. Such effort up front avoids many problems with individual task elements as well as often avoiding the problem of one incomplete task preventing the successful completion of other tasks. The Pan-STARRS management and engineering team has invested considerable effort systematically developing the PS2 commissioning plan, identifying critical and/or risk points, developing contingency plans, and lowering risk as much as possible before integration and test actually begins.

Primary factors increasing risk for the three basic assessment categories for an integration and test program are as follows:
- Budget
    - Technical component or features not mature so require development or repair beyond that planned,
    - Poor schedule estimates,
    - Unknown costs, i.e., no work element with associated costs for material, fabrication, equipment, or personnel.
- Schedule
    - Unknown tasks,
    - Poor estimate of duration.
- Technical
    - Poor design,
    - Initial design validation revealing problems,
    - System HW does not perform properly,
    - Lack of tools to test and analyze performance,
    - Requirements/scope creep.

Details of specific risk analysis cases for PS2 are provided in Section 3.3.

**2.3 Primary lessons learned from PS1**

Experiencing and learning from all of the PS1 growing pains is exactly what was intended when the long term management model was implemented at the beginning of the project in 2003 and 2004: PS1 was specifically built as a prototype and demonstration system before investing in the follow-on augmentations with additional telescopes and cameras (PS2 through the completion of PS-4). This was done in order to demonstrate the validity of the design, identify fabrication issues, complete the development tasks which the IfA/Pan-STARRS team decided to do internally, and learn how to commission and then operate the system. Note that having this management model as an option is a distinct strength of a distributed aperture system, and the augmentation of PS1 with PS2 strongly leverages that experience. The approach was not only developed internally by the Pan-STARRS team, but was also strongly endorsed by a succession of external advisory committees convened to review, assess, and recommend the management and engineering approach to developing Pan-STARRS. Membership on those committees included at one time or another experts such as Dr. James Beletic, Dr. Donald Yeomans, Dr. Steve Shectman, Dr. Roc Cutri, Dr. Scott Tremaine, Dr. Simon P. Worden, Dr. Robert Fugate, Mr. Paul Kervin, and Mr. Robert Hunt.

The single most important lesson learned from PS1 is that, in fact, the Pan-STARRS design is fundamentally sound and delivers very close to the expected theoretical performance, and that while challenging in many respects, it is possible to fabricate the design to, or nearly to, design specifications. This is often not sufficiently appreciated, especially for the optics. The limiting factors for PS1 performance are fabrication issues, and how lessons learned from PS1 fabrications of the telescope (including optics) and camera (including CCDs) are incorporated in the PS2/GPC2 fabrications is addressed in more detail in Section 3.3 of this paper as well as the papers by Morgan et al. and Onaka et al. presented at this conference (op. cit.).

The second critical lesson learned from PS1 is to, as much as possible, complete fabrication and development before proceeding to commissioning. Not doing so significantly increases schedule risk, especially when many or all of a team are responsible for both development and commissioning tasks as has been the case with Pan-STARRS. For PS2, mitigating this problem has been a specific focus of the management plan, and has included some increased staffing and increased lead times.

The third primary lesson learned from PS1 is quantifying how well PS1 actually performs, and how the performance can be improved for PS2. These improvements are summarized in Section 3.1.

# 3. PS2 AS OF JULY 2012

## 3.1 Improvements of PS2 relative to PS1

The final milestone goal for this phase of Pan-STARRS is to augment and upgrade the distributed aperture Pan-STARRS with an additional telescope, and then integrate and test the augmented system to bring it to operational readiness in 2014. The new telescope and camera are individually referred to as PS2 and GPC2, respectively (or, depending on the context, sometimes PS2 will refer to the integrated telescope and camera system), and the augmented Pan-STARRS is often referred to as PS1+2. As of July 2012, the remaining scope of work for the PS2 phase consists of the necessary installation, integration, and testing work needed to bring the PS1+2 summit system to operational readiness; there are no other tasks requiring funding support or time from other sources.

At this point, almost all of the equipment or fabrication costs required to perform the statement of work have already been paid or funding resources exist. Specifically, the optics, filters, camera CCDs ("detectors"), camera electronics, camera materials for the cryostat, camera shutter, and the telescope structures have been completed or are all fully funded. In this context, "telescope structures" refers to the main mount and structure being fabricated by Advanced Mechanical and Optical Systems (AMOS) in Belgium, the filter mechanism, the "Upper Cassegrain Core" (UCC) that supports two of the three corrector lenses, and the "Lower Cassegrain Core " (LCC) that supports the instrument package composed of the filter mechanism, camera shutter, and GPC2 camera.

As of July 2012, all of the telescope, camera, and OTIS design work is complete. The polishing of all of the optics has been completed, the three corrector lenses have been coated with anti-reflection (AR) coatings, the camera shutter has been delivered, all six filters have been delivered, the filter mechanism is in the final stages of integration, camera machining tasks are ~80% complete, and camera electronics and device integration begins in September. By the end of July 2012, both the LCC and UCC will also be complete (including the integration of the L1 and L2 corrector lenses into the UCC). Approximately half of the CCDs required for GPC2 will be delivered, tested, and integrated into GPC2 before the end of 2012; the remaining devices required to completely populate the 64-device GPC2 focal plane will be delivered in early 2013. The CCDs are being fabricated under a contract with the Massachusetts Institute of Technology Lincoln Laboratory (MITLL). The telescope structure being fabricated by AMOS is scheduled to be delivered to the observatory site in early 2013.

It is worth emphasizing there is strong evidence from design improvements and the as-builts that PS2 will be better than PS1 in terms of both performance and reliability in these areas:
- Better telescope structure,
- Better optics,
- Better filters,
- Better CCDs,
- Better camera electronics (reduced cross talk, reduced read noise, improved thermal control),
- Significant reduction of ghosting and scattered light,
- Mature image processing due to the efforts of the SW engineers belonging to the PS1 Science Consortium.

These improvements also serve as early risk mitigation against some of the adverse impacts encountered during PS1.

## 3.2 PS2 current status and work plan

A comprehensive integrated project schedule (IPS) has been generated with an associated work breakdown structure (WBS) to Level 5. The WBS has been created from a full bottom-up resource-loaded analysis of the tasks required to complete the proposed scope of work (SOW). We use this WBS and IPS to support the estimates of required manpower, schedule and budget, and risk assessments. The work plan methodology includes substantially leveraging the design, technology, and techniques that proved effective during the fabrication of PS1. Significant "Lessons Learned" have been incorporated to improve the telescope, camera, and OTIS subsystems, and to increase the efficiency of the commissioning efforts.

Major milestones or work blocks and the estimated schedule as of July 2012 include:
- The On-Site Delivery (OSD) of the telescope structure (before February 2013),
- Completion of optics integration with the telescope structure (before May 2013),
- The delivery to the site of the camera with ~40-50% of final number of CCDs, "GPC2-25" (February 2013),
- Completion of the instrument integration with the telescope, including GPC2-25 (in April 2013),

- The Site Acceptance Testing (SAT) of the telescope (completed May-June 2013),
- The start of post-SAT telescope commissioning, i.e., testing that does not involve direct participation by the telescope vendor (June 2013),
- The return of GPC2-25 to Oahu for integration with the remaining CCDs, "GPC2-64" or simply GPC2, and focal plane metrology to match the physical detector surface to the actual optical focal surface determined from the preliminary collimation and alignment (September 2013)
- Subsequent return of GPC2 to the summit (October 2013),
- Final collimation and alignment (October-November 2013),
- The successful simultaneous operation of the PS1 and PS2 telescopes using OTIS (November-December 2013).

The telescope-related activity during this performance period will be assembly of the telescope components during on-site integration, the acquisition of site acceptance test (SAT) data, the analysis of SAT data, and the subsequent PS2 on-site commissioning. After SAT has been completed, the responsibilities of the telescope vendor are complete, and the Project then proceeds with commissioning of the system for science use. The principal task is then collimation and alignment (C&A) to optimize image quality. Due to the extensive experience with PS1 C&A, and the development of the techniques and necessary tools that are now mature, it is possible to accurately predict the effort required to complete this critical work.

The period between Telescope On-Site Delivery (OSD) and SAT requires the most time and labor intensive efforts for the Project on-site personnel. To save time in the schedule, we plan on bonding the support fixtures on M1 much earlier than the telescope OSD, prior to the coating of M1. However, this is not possible with M2. The on-site telescope integration will therefore begin with the important step of bonding the M2 support fixtures. The on-site integration will then continue with the placement of the mirrors into their cells and the corrector lens support structure (the UCC) into the telescope. The installation of the telescope into the dome includes the lifts required to place the telescope base, fork, center section, truss, and secondary supports into the dome. These efforts will be done in conjunction with AMOS personnel. Between the installation of the telescope in the dome and the installation of the Filter Mechanism (FM) are several tasks related to verification of the telescope metrology which will be done by Project personnel. These tasks include the installation of the LCC and UCC support structures along with their associated corrector lenses. Initial alignment of these structures will be done by Project personnel. The baffle and temperature sensor installation, range of motion, limit switch tests, following-error step and settle tests, and all site acceptance testing will be done in conjunction with telescope vendor personnel. Compared to PS1, the Factory Acceptance Testing (FAT) that occurs in 2012 is much more extensive and will validate many critical telescope performance features prior to OSD, and much of the SAT work will simply repeat work done for FAT.

Since the project will re-use the operational GPC1 camera software deployed for PS1, the camera control software effort (GPC2 Software Development) is largely complete except for the creation of new configuration parameter files specific to the CCDs actually used in GPC2. The installation of the GPC2 computers and on-site network computers includes work to aggregate the Gigabit fiber links into 10-Gigabit links. Other work blocks for GPC2 include:
- OTA device testing,
- Focal plane thermal control,
- Upgrades to the observatory interfaces.

The telescope and observatory control software (OTIS) used for PS1 requires some upgrades to more efficiently control the individual observatory/telescope/camera subsystems, and new capability to implement the option to operate PS1 and PS2 synchronously. The synchronous mode enables forming a summed image by combining effectively simultaneous multiple images to maximize detection sensitivity and limit any apparent PSF smearing. This is a unique capability for PS1+2 intrinsic to the distributed aperture concept. This is particularly important for NEO searches when using PS1+2 to observe the same patch of sky. In addition to the basic operator-level user software, the low level telescope control system for PS2 and the survey-specific scheduling software are considered to be OTIS components. The PS1+2 user-level software and scheduling tools are strongly leveraged by the existing PS1 versions.

### 3.3 PS2 risk assessment

Beginning from the conclusion of PS1 commissioning, the Pan-STARRS management and engineering team has invested considerable effort systematically developing the PS2 commissioning plan, identifying critical risk points, developing contingency plans, and lowering risk as much as possible before summit integration and test actually begins. This provides the basis for our conclusions that the cumulative risk for the proposed work is generally Low. This includes, e.g., a Low Technical risk for the overall system concept and performance because the designs are mature and validated by PS1 system performance plus the known characteristics of the PS2 as-builts.

As noted previously, the risk assessment was carried out beginning from the comprehensive task list in the integrated Project schedule that goes to five levels in many cases. We believe there is no additional risk to completing commissioning of the telescope, camera, and OTIS due to tasks not included in the IPS/WBS as we believe the essential tasks have been incorporated into the current WBS. However, the status and path forward are reviewed regularly and often. The WBS identifies a large number of tasks necessary to execute the work prior to and during the performance period. The amount of effort expended to define the tasks is deliberate as we have found that in practice, the risk is reduced by such detail: as the schedule and budget estimates increase their fidelity, it becomes easier to identify potential technical issues. Then, as much as possible, risk control can occur in advance by identifying, preparing, and implementing potential mitigation strategies as necessary.

It is well understood and appreciated that it is certainly possible that a sequence of low risk tasks (low consequence, low probability of occurrence) might pose a cumulative risk of greater severity, but that is not automatically true. It is also possible that the cumulative risk remains low. Should an unfortunate sequence of lower probability low consequence events actually occur, we have attempted to explicitly accommodate mitigation within the proposed contingency and other protections such as the telescope warranty and quick access to replacement parts. We do believe that our understanding of the required tasks combined with our experience of successfully completing these tasks once before for PS1 explicitly reduces the risk for successfully completing them on budget and schedule for PS2: this is standard methodology, not "spin". However, as we conclude in the final paragraph of this section, we do not claim to be perfect and all-knowing, and we fully expect to encounter some surprises along the way that will require some additional contingency and/or time beyond what we have already budgeted and scheduled.

In general, the Technical and Schedule risks for PS2 are simply not the same as for PS1. One very important reason that is true is that the same team that designed and subsequently commissioned PS1 designed and will commission PS2; in other words, successfully completing PS1, with PS1 operated successfully and reliably since, constitutes the single biggest risk reduction and mitigation factor for PS2. Consider that the PS1 "experience" demonstrated that

- The overall PS1 design was validated and is sound (it is not always sufficiently appreciated or is forgotten that so many aspects of the PS1 design were completely new, and yet PS1 works as designed for the most part),
- The design is tightly coupled to the defined capability requirements,
- The deviations in PS1 performance from the capability requirements and goals are mostly due to fabrication issues, not design issues.

Given that, the Pan-STARRS team has focused a great deal of effort since 2009 on correcting and better controlling the fabrication issues encountered during PS1. From the PS2 as-builts, we already know those efforts have been quite successful (e.g., in the optics, filter, and CCD fabrications).

Furthermore, the PS1 commissioning plan was a first time effort in many respects due to the unique nature of PS1, but those lessons have also been learned very well and not forgotten in the planning and execution of PS2. One of the most important lessons learned was the importance of completing as much development as possible before the dedicated testing begins (for both HW and SW). The HW development for all telescope-related tasks will be complete before the telescope OSD. While some new SW development is occurring for PS2, most of the core control, data reduction, and data analysis SW is very similar or identical to what was used for PS1. Of particular note is that the extensive tool kit of SW required specifically for commissioning already exists for PS2, e.g., the SW doing the collimation and alignment calculations to optimize the optical performance. With these validated tools already in hand, planning and executing the commissioning plan becomes more systematic, more clearly defined from start to finish, and better controlled, i.e., lower risk.

The Budget risk for non-catastrophic events is Low with respect to the proposed SOW as long the actual durations for the commissioning tasks do not deviate too much from the estimated durations because the fabrication costs have been paid or encumbered (fully committed from existing funds). The actual durations from PS1 commissioning have been used as inputs into the schedule estimates for the associated WBS elements, and given that by the end of PS1 commissioning, the estimated durations were usually quite close to the estimates, we have some confidence in the initial PS2 estimates. The Schedule risk for the telescope and associated testing is Low through the Site Acceptance Testing because almost all of the long lead fabrication will be complete by January 1, 2013, but with minor delays still possible in the final delivery of the AMOS telescope structure or the MITLL CCDs. Careful and frequent monitoring and interaction with the vendors is occurring to identify schedule risks at a finer level of detail (and be mitigated as necessary). For relatively minor additional fabrication, we have set aside contingency for additional machine shop hours, engineering, and construction material. Schedule contingency of 75 days for weather, installation, integration, and commissioning delays has been explicitly added to the Project schedule, and a substantial schedule contingency also has been included for SW tasks. The weather contingency is based on 10-year historical Haleakala meteorological data.

In particular, note that the problems experienced during PS1 commissioning due to issues with the telescope have been explicitly mitigated through the selection of a new telescope vendor (AMOS) providing a more robust telescope. Important improvements to the PS2 telescope relative to that of PS1 that act as direct risk mitigation for and during commissioning (and subsequent operational performance) include the following:
- Actively cooled drives with increased performance margin,
- Cooled primary mirror assembly (PMA)
- Better M2 support,
- A stiffer truss providing increased margin against vibration,
- Better M1 figure control utilizing 36 active figure controllers,
- Better overall reliability not only due to intrinsic design but better quality control of as-built components.

As noted, other potential technical risks related to telescope issues are partially to substantially mitigated because the telescope will be under warranty well past the commissioning period (unlike the PS1 telescope).

Due to minor telescope fabrication delays over the last 14 months at AMOS, the telescope OSD date was initially assessed as a Medium risk. However, we believe the actual risk is Low since FAT is now only five months away with no further delays anticipated by AMOS, and due to specific measures implemented by AMOS to control the remaining schedule to Factory Acceptance Testing (the last step before shipping PS2 to Maui).

The CCD device delivery is an area requiring special attention and mitigation due to delays by MITLL. The MITLL fabrication consists of two phased production runs, assembly of these CCDs onto their mechanical packages and subsequent testing by the Project. To mitigate the impacts of delayed detector deliveries, we are planning to do an initial assembly of the GPC2 camera with less than 64 of the devices from the first production phase populating the focal plane. In this manner, we have a conservative path forward that allows us to do site acceptance testing of the telescope while the remainder of the camera devices are delivered and tested. This intermediate step also provides the opportunity to make final adjustments to both the detector voltage settings and to the device placements on the focal plane after acquiring on-sky information on the optical performance of the telescope. This step was proven to be extremely useful in the development of PS1. At the moment, the Project expects to begin on-sky observing using a less than fully populated focal plane camera denoted as "GPC2-25" for the expected number of ~25 devices to be live on the focal plane. However, as we assess the Schedule risk for the GPC2-25 delivery to be Medium, we have developed two contingency plans to remove this from the critical path and keep the overall Schedule risk at Low for completing SAT: (1) we can use an existing test camera containing 16 OTA detectors in place of GPC2-25 with no impact on SAT or even the first stages of post-SAT collimation and alignment, or (2) if the test camera could not be used for some reason that we cannot really envision, we could still complete SAT with a COTS video camera. It is also important to note again that in spite of the CCD fabrication delays, MITLL has remained committed to providing, and is achieving, devices representing a significant upgrade to the devices used in GPC1.

In terms of the risk for completing OTIS software development, we mention two work blocks in particular: the telescope control SW (TCS) for which the Project has responsibility, and the scheduling tools required to create nightly and monthly observing programs and track survey completion metrics. Much of the TCS development is new, and could have been assessed at least as Medium risk. To mitigate that in advance, we have purchased validated SW used successfully with other telescopes, viz., the TCSpk package available from TPoint Software and the TPK package available from the Science Technology and Facilities Council. TCSpk provides validated routines for all components of the TCS related to, e.g., telescope axis control, track computation, and pointing, and also includes the full SLALIB library routines. The TPK package provides an overall object-oriented framework in which TCSpk can be embedded. This then provides a natural integration path within the higher-level OTIS architecture that eventually ends with the user-interface by which the telescope "observers" operate the system. This reduces the amount of actual development to only needing to integrate the building blocks as opposed to creating all the code from scratch. In addition, we have contacted SW engineers with other projects who have used these tools, and who are willing to consult on an as-needed basis, and the two vendors for TCSpk and TPK have also offered consulting help if necessary. Given the maturity and validation history of the packages combined with a deliberately large schedule contingency for development time and the plan to test these OTIS components at FAT, we believe the risk is fairly assessed as Low for implementing this capability beginning with post-FAT commissioning.

The development of the scheduling tools for PS1+2 observing is strongly based on those that already exist for PS1. Improvements to some of the existing tools plus some new capabilities and features are currently planned. The senior SW engineer responsible for writing all of the PS1 scheduling SW and much of the MOPS SW for asteroid detections will also develop the PS1+2 scheduling SW. The continuity of an experienced, highly expert engineer is crucial to lowering the risk for having these tools completed before the beginning of PS1+2 operations. Thus, the Technical Risk is Low, and there is no Budget risk applicable other than that associated with work time. As the current schedule contains contingency with a 2-3x time factor, both the Schedule and associated Budget risk are Low.

Finally, we note that while we are confident that the PS2 commissioning will proceed much more efficiently than for PS1, we are not over-confident. We fully recognize that unanticipated issues will arise, and some of those may incur delays or added expense. The PS1 and PS2 observatories are shown in Figure 3 for reference.

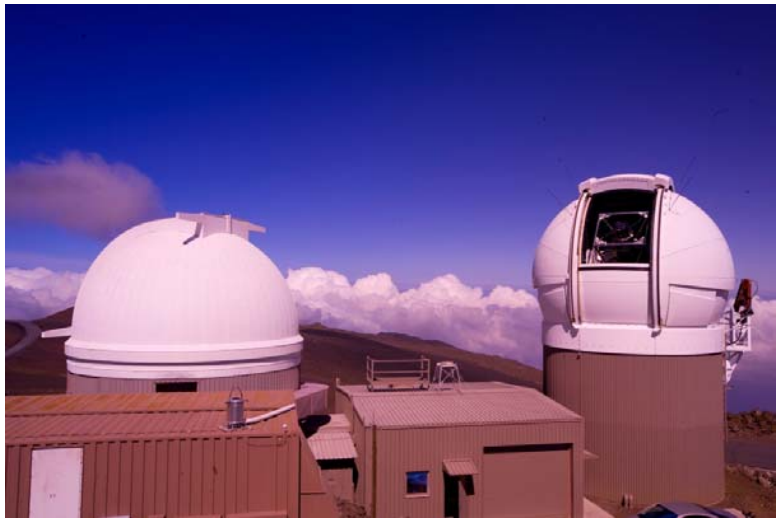

Figure 3. On the right is the PS1 observatory, and on the left is the future home of PS2.